\documentclass[
aps,
 ,twocolumn
]{revtex4-1}

\usepackage{times}
\DeclareMathAlphabet{\mathcal}{OMS}{cmsy}{m}{n}

\usepackage{graphicx}
\usepackage{amsmath}
\usepackage{enumitem}
\usepackage{amssymb}
\usepackage{array}
\newcolumntype{L}[1]{>{\raggedright\let\newline\\\arraybackslash\hspace{0pt}}m{#1}}
\newcolumntype{C}[1]{>{\centering\let\newline\\\arraybackslash\hspace{0pt}}m{#1}}
\newcolumntype{R}[1]{>{\raggedleft\let\newline\\\arraybackslash\hspace{0pt}}m{#1}}
\usepackage{mathrsfs}
\usepackage{dsfont}
\usepackage{appendix}
\usepackage{multirow}
\usepackage{tikz}
\usepackage{pgflibraryarrows}
\usepackage{pgflibrarysnakes}
\usetikzlibrary{decorations.pathmorphing}
\usepgflibrary{arrows.meta}

\usepackage{subfigure}

\DeclareMathAlphabet{\mathcal}{OMS}{cmsy}{m}{n}

\newcommand{\D}{\mathrm{d}}

\newcommand{\dd}{\dagger}
\newcommand{\al}{|\alpha|^2}

\usepackage{braket}

\definecolor{blueprl}{RGB}{46,48,146}
\usepackage[pdftex,colorlinks=true,urlcolor=blueprl,citecolor=blueprl,linkcolor=blueprl]{hyperref}
\newcommand{\vt}{\vphantom{\frac{1}{\sqrt{2}}}}
\newcommand{\non}{\nonumber}

\begin{document}

\title{Continuous-variable quantum teleportation with vacuum-entangled Rindler modes}

\author{Joshua Foo}
\email{joshua.foo@uqconnect.edu.au}
\author{Timothy.~C.~Ralph}\email{ralph@physics.uq.edu.au}
\affiliation{Centre for Quantum Computation and Communication Technology, School of Mathematics and Physics, The University of Queensland, St. Lucia, Queensland, 4072, Australia}

\date{\today}

\begin{abstract}
{We consider a continuous-variable quantum teleportation protocol between a uniformly accelerated sender in the right Rindler wedge, a conformal receiver restricted to the future light cone, and an inertial observer in the Minkowski vacuum. Using a nonperturbative quantum circuit model, the accelerated observer interacts unitarily with the Rindler modes of the field, thereby accessing entanglement of the vacuum as a resource. We find that a Rindler-displaced Minkowski vacuum state prepared and teleported by the accelerated observer appears mixed according to the inertial observer, despite a reduction of the quadrature variances below classical limits. This is a surprising result, since the same state transmitted directly from the accelerated observer appears as a pure coherent state to the inertial observer. The decoherence of the state is caused by an interplay of opposing effects as the acceleration increases: the reduction of vacuum noise in the output state for a stronger entanglement resource, constrained by the amplification of thermal noise due to the presence of Unruh radiation. 
}
\end{abstract}
\maketitle
\section{Introduction}
Entanglement is a fundamental property of the vacuum state of relativistic quantum field theory. It arises when spacetime is partitioned into distinct regions and appears between the resulting subsystems \cite{birrell1984quantum}. For example, in flat Minkowski spacetime, the vacuum state can be expressed as an entangled state between the complete sets of modes spanning the left and right Rindler wedges. This gives rise to the Unruh effect, which predicts that a uniformly accelerated detector sees a thermal bath of particles. Since the detector is confined to the right Rindler wedge, the unobserved Rindler modes in the left wedge are traced out, yielding a thermalised vacuum state \cite{unruh1976notes,crispino2008unruh,alsing2004simplified,davies1975scalar}. 

The extraction of this underlying vacuum entanglement has been well studied in the literature. The seminal work of Reznik et al. \cite{reznik2003entanglement} established the field of \textit{entanglement harvesting}, which considers the swapping of vacuum entanglement onto bi-partite quantum systems, such as Unruh-deWitt detectors, through local interactions. Such harvesting protocols have been applied to spacelike \cite{reznik2005violating} and timelike \cite{olson2011entanglement,olson2012extraction} separated detectors, as well as situations involving uniform accelerations \cite{salton2015acceleration}, black holes \cite{ng2018unruh,henderson2018harvesting} and expanding universes \cite{nambu2013entanglement,kukita2017entanglement,martin2014entanglement,ver2009entangling}.

A natural question to consider is whether the intrinsic entanglement of spacetime can be utilised as an information-theoretic and physical resource for quantum communication protocols. Previous work by Ralph et al. uses observers coupling to the vacuum-entangled modes of a massless scalar field (by operating detectors whose time-dependent energy scaling imitates a uniformly accelerated observer interacting with the Rindler modes \cite{olson2011entanglement,olson2012extraction,ralph2015quantum}) to implement a quantum key distribution protocol, whilst Reznik has offered proof-of-principle proposals for dense coding and quantum teleportation protocols through vacuum entanglement swapped onto pairs of stationary, spacelike-separated atoms \cite{reznik2000distillation}. In \cite{koga2018quantum}, the authors employ two Unruh-deWitt detectors in relative inertial motion that interact perturbatively with the field to extract entanglement and perform better-than-classical teleportation. 

The focus of this paper is also vacuum entanglement-enabled quantum teleportation, however here, we adopt a \textit{nonperturbative, continuous-variable approach}, formulated with observers in \textit{relativistic, non-inertial reference frames}. Whilst teleportation protocols with accelerated partners have been previously studied \cite{landulfo2009sudden,lin2015quantum,alsing2003teleportation,alsing2004teleportation}, they generally feature qubit degrees of freedom (Bell state measurements and two-level detectors) and a pre-existent entanglement resource (such as entangling photon cavities) which is distributed between the observers. The main results from those papers is that the Unruh radiation perceived by the accelerated observer typically leads to the degradation of the fidelity of the protocol.  

Here, we consider a continuous-variable teleportation protocol in which Rob, who uniformly accelerates in the right Rindler wedge, uses the entanglement of the quantum vacuum to teleport unknown Rindler-displaced Minkowski vacua (displaced thermal states) to a stationary `conformal receiver', Charlie (who resides in the same reference frame as Rob, by virtue of being restricted to the future light-cone of Minkowski spacetime), who then directly transmits the teleported state to Alice, an inertial observer in the Minkowski vacuum. We employ the nonperturbative quantum circuit model developed in \cite{su2017quantum, su2019decoherence} to describe the evolution of the initial state between the observers. Our main result indicates that although performance beyond the classical channel limit is possible, the limit of high entanglement does not lead to ideal, continuous-variable teleportation in this scenario.

This paper is organised as follows: in Sec.\ \ref{II}, we review the continuous-variable teleportation protocol which we will implement in the reference frame of the uniformly accelerated observer. In Sec.\ \ref{sec:III}, we discuss the notion of vacuum entanglement between independent regions of spacetime, before reviewing the quantum circuit model and the self-homodyne detection technique for the inertial observer in Sec.\ \ref{sec:IV}. In Sec.\ \ref{sec:V}, we present analytic results for the purity of output states according to the inertial Minkowski observer. We conclude with some final remarks and opportunities for further research in Sec.\ \ref{sec:VI}. 

\section{Continuous-Variable Teleportation Between Inertial Observers}\label{II}
Quantum teleportation describes the transfer of an unknown quantum state $\hat{\rho}_\mathrm{in}$ between distant observers using a classical channel and a preexisting entanglement resource \cite{bennett1993teleporting}. Whilst originally formulated using entangled Bell pairs as the resource, a continuous-variable version of the protocol was first introduced by Vaidman in \cite{vaidman1994teleportation} and then by Braunstein and Kimble in \cite{braunstein1998teleportation}. We follow the all-optical approach in \cite{ralph1999all}, where the classical channel is enacted via the direct transmission of a classical field wherein the conjugate quadrature operators of the field are largely amplified above the quantum noise limit (QNL). The quantum channel involves the distribution of an entangled bi-partite system between the observers. By performing appropriate local operations on their entangled system, the receiver can retrieve an arbitrarily good version of the initial state $\hat{\rho}_\mathrm{in}$ without any quantum information being directly transmitted. For our purposes, the utility of this approach is that the evolution of field modes can be modeled as unitary operations. 

Consider the diagram shown in Fig.\ 1. 
\begin{figure}[h]
    \centering
    \includegraphics[width=0.7\linewidth]{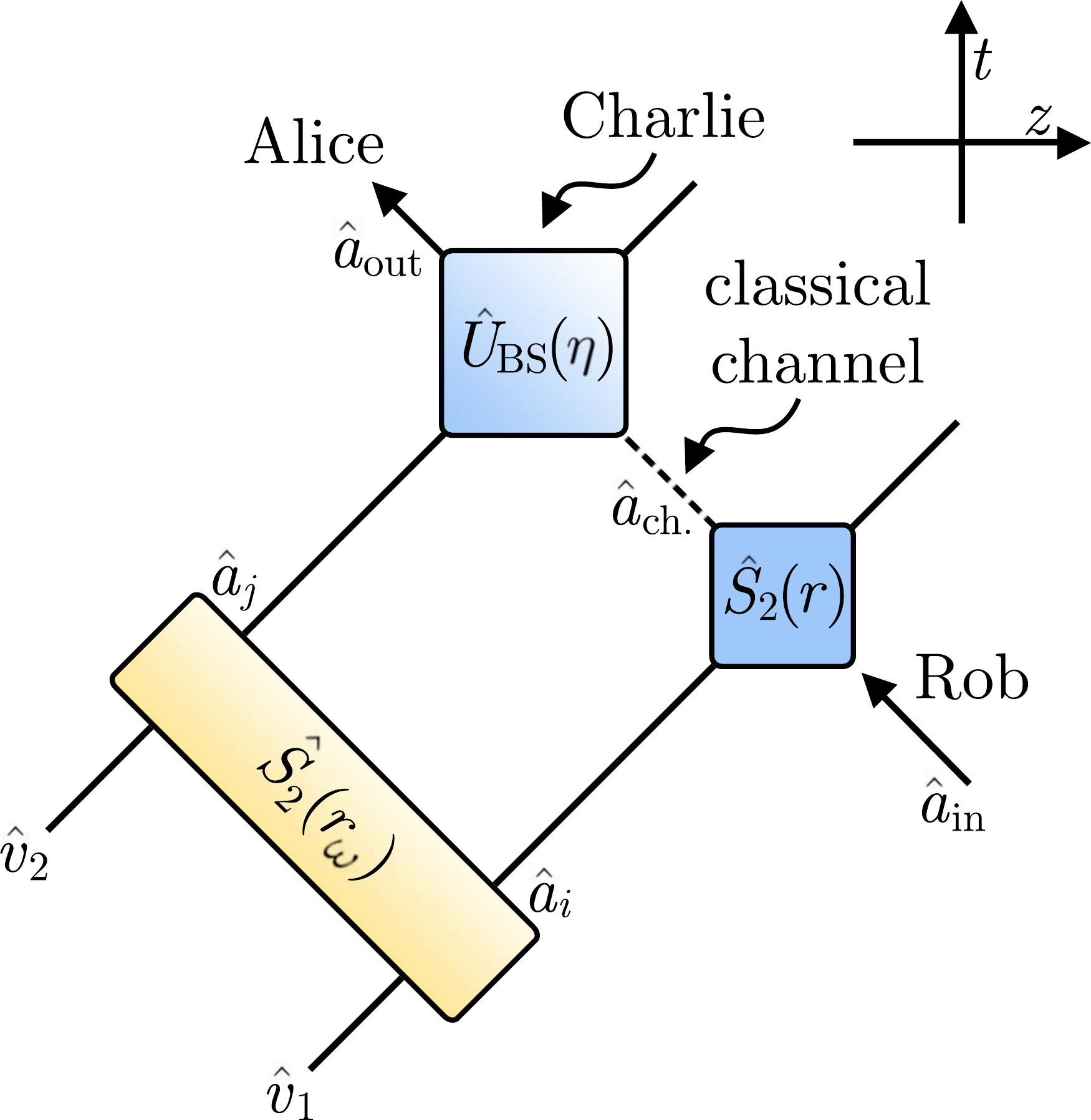}
    \caption{The all-optical teleportation protocol between inertial observers. Rob prepares an unknown state in the mode $\hat{a}_\mathrm{in}$ and mixes it at a linear optical amplifier (modeled by a two-mode squeezing unitary) with one half of an EPR-entangled field mode, $\hat{a}_i$. The amplification of the input state generates an effective classical channel between Rob and Charlie, who combines $\hat{a}_\mathrm{ch.}$ with the other half of the EPR pair at a beam splitter. If Charlie selects $\eta$ appropriately, Alice can receive a perfect reconstruction of $\hat{a}_\mathrm{in}$, in the limit of high entanglement.}
    \label{fig:my_label}
\end{figure}
A pair of EPR-entangled field modes is distributed between two distant observers, Rob and Charlie. Rob, operating a linear optical amplifier, mixes the input mode $\hat{a}_\mathrm{in}$ with the entangled vacuum mode $\hat{a}_i$ and transmits the output field $\hat{a}_\mathrm{ch.}$ to Charlie. For a sufficiently large gain $\mathcal{G}$, the conjugate quadrature operators $\hat{X}_\mathrm{ch.}^+, \hat{X}_\mathrm{ch.}^-$ have uncertainties much larger than the quantum limit of unity. Hence, the quantum noise introduced by a joint measurement of $\hat{X}_\mathrm{ch.}^+, \hat{X}_\mathrm{ch.}^-$ is negligible compared to the already amplified quadrature amplitudes, allowing for the designation of $\hat{a}_\mathrm{ch.}$ as a classical field \cite{ralph1999all}. Charlie receives $\hat{a}_\mathrm{ch.}$ and mixes it at a beam splitter with the other half of the entanglement resource, $\hat{a}_j$. By attenuating $\hat{a}_\mathrm{ch.}$ by $1/\mathcal{G}$ and then considering the limit of perfect entanglement between the vacuum modes $\hat{a}_i$ and $\hat{a}_j$, Charlie is able to retrieve the input mode $\hat{a}_\mathrm{in}$. 

To illustrate this mathematically, consider the evolution of the bosonic operator $\hat{a}_\mathrm{in}$ through the teleportation protocol, given in the Heisenberg picture by
\begin{align}
    \hat{a}_\mathrm{out} &=  \hat{S}_2^\dd (r) \hat{U}_\mathrm{BS}^\dd(\eta) \hat{a}_\mathrm{in} \hat{U}_\mathrm{BS} (\eta)  \hat{S}_2(r),
    \vt 
\end{align}
where
\begin{align}
    \hat{S}_2(r) &= \exp \Big[\xi^\star \hat{a}_i\hat{a}_\mathrm{in} - \xi  \hat{a}_\mathrm{in}^\dd \hat{a}_i^\dd  \Big] ,
    \vt 
    \\
    \hat{U}_\mathrm{BS}(\eta) &= \exp \Big[ i\theta\left( e^{i\phi} \hat{a}_\mathrm{in}^\dd  \hat{a}_j + e^{-i\phi}\hat{a}_\mathrm{in} \hat{a}_j^\dd \right) \Big] .
    \vt 
\end{align}
$\hat{S}_2(r)$ is a unitary two-mode squeezing operator with amplitude $\xi = re^{i\alpha}$, describing the linear amplification of $\hat{a}_\mathrm{in}$ mixed with $\hat{a}_i$, whilst $\hat{U}_\mathrm{BS}(\eta)$ represents the beam splitter interaction performed by Charlie. Recalling that in the Heisenberg picture, operators act on the modes in reversed temporal order, we have
\begin{align}
    \hat{a}_\mathrm{out} &= \hat{S}_2^\dd (r) \big(\sqrt{\eta} \hat{a}_\mathrm{in} - \sqrt{1 - \eta} \hat{a}_j  \big) \hat{S}_2(r) ,
    \vt 
    \non 
    \\
    &= \sqrt{\eta} \hat{S}_2^\dd(r) \hat{a}_\mathrm{in} \hat{S}_2(r)  - \sqrt{1 - \eta} \hat{a}_j ,
\end{align}
where we have used the substitution $\eta = \cos^2\theta$. Now, $\hat{S}_2^\dd(r) \hat{a}_\mathrm{in} \hat{S}_2(r)$ imitates the linear amplification of the field quadratures of $\hat{a}_\mathrm{in}$ mixed with $\hat{a}_i$ for $r\gg 1$, acting as the classical channel between Rob and Charlie,
\begin{align}
    \hat{a}_\mathrm{out} &= \underbrace{ \sqrt{\eta} \cosh (r) \hat{a}_\mathrm{in} + \sqrt{\eta}\sinh (r) \hat{a}_i^\dd }_{\hat{a}_\mathrm{ch.}} - \sqrt{1 - \eta} \hat{a}_j .
    \vt 
\end{align}
It can also be regarded as a joint, continuous-variable measurement of the quadrature operators for $\hat{a}_\mathrm{in}$ and $\hat{a}_i$. Then, by taking the attenuation of Charlie's beam splitter to be $\eta = (\cosh r )^{-2}$, $\hat{a}_\mathrm{out}$ reduces to
\begin{align}\label{eq:tanh}
    \hat{a}_\mathrm{out} &= \hat{a}_\mathrm{in} + \hat{a}_i^\dd \tanh (r) - \hat{a}_j \sqrt{1 -  \frac{1}{\cosh^2(r)} }.
\vt 
\end{align}
In the limit $r\gg 1$, Eq.\ (\ref{eq:tanh}) becomes 
\begin{align}\label{eqn:6}
    \hat{a}_\mathrm{out} &\simeq \hat{a}_\mathrm{in} + \hat{a}_i^\dd - \hat{a}_j,
    \vt 
\end{align}
implying that the output state is polluted by two additional QNL from $\hat{a}_i, \hat{a}_j$. Recall however that $\hat{a}_i$ and $\hat{a}_j$ are highly correlated via two-mode squeezing of the vacuum state, related by
\begin{align}
    \hat{a}_i &= \hat{S}_2^\dd (r_\omega) \hat{v}_1 \hat{S}_2(r_\omega) = \hat{v}_1\cosh (r_\omega) + \hat{v}_2^\dd \sinh (r_\omega) , 
    \vt 
    \\
    \hat{a}_j &= \hat{S}_2^\dd(r_\omega) \hat{v}_2\hat{S}_2(r_\omega) =  \hat{v}_2 \cosh (r_\omega) + \hat{v}_1^\dd  \sinh (r_\omega) ,
    \vt 
\end{align}
where $r_\omega$ is the squeezing amplitude. Expressing $\hat{a}_i^\dd$ and $\hat{a}_j$ in terms of $\hat{v}_1,\hat{v}_2$ yields
\begin{align}\label{eqn:9}
    \hat{a}_\mathrm{out} &= \hat{a}_\mathrm{in} + \left( \cosh r_\omega - \sinh r_\omega \right) \big( \hat{v}_1^\dd - \hat{v}_2 \big).
    \vt 
\end{align}
In the limit of perfect entanglement between $\hat{a}_i$, $\hat{a}_j$ ($r_\omega \to \infty)$, then
\begin{align}
    \hat{a}_\mathrm{out} &= \hat{a}_\mathrm{in} .
    \vt 
\end{align}
In this limit, the effective channel between the input and output is the identity. This protocol can be classified as quantum teleportation because the only direct link between the input and output is the classical channel through which the highly amplified field propagates \cite{ralph1999all}. 
    
\section{Vacuum Entanglement in (1+1)-Dimensional Quantum Field Theory}\label{sec:III}
We now review the entanglement structure of the vacuum state in (1+1)-dimensional quantum field theory.
Consider a massless, scalar field $\hat{\Phi}$ which can be expanded in a plane wave basis, given by
\begin{align}\label{eqn:minkowskifield}
    \hat{\Phi} (U,V) &= \int \D k \Big( \hat{a}_{kl} u_{k}(V) + \hat{a}_{kr} u_{k}(U) + \mathrm{h.c} \Big),
    \vt 
\end{align}
where h.c denotes the Hermitian conjugate, and $u_{k}(V) (u_{k}(U))$ are left-moving (right-moving) single-frequency mode functions \cite{su2017quantum}
\begin{align}
    u_{k} (V) &= \frac{1}{\sqrt{4\pi k}}e^{-ikV} ,
    \vt 
    \\
    u_{k} (U) &= \frac{1}{\sqrt{4\pi k }} e^{-ikU},
    \vt 
\end{align}
where $V = t+z$ and $U= t-z$ are Minkowski light-cone coordinates and $k$ is the frequency of the mode with respect to $t$. The single-frequency Minkowski operators $\smash[b]{\hat{a}_{kl} (\hat{a}_{kr})}$ and $\smash[b]{\hat{a}_{kl}^{\dd} (\hat{a}_{kr}^{\dd} )}$ satisfy standard bosonic commutation relations, $\smash[b]{[\hat{a}_{ki},a_{k'i'}^{\dd}] = \delta_{ii'}\delta(k-k')}$ and $\smash[b]{[\hat{a}_{ki},\hat{a}_{k'i'}] = 0}$, where $i= l,r$ denotes the directionality. The single-frequency Unruh operators $\hat{c}_\omega$, $\hat{d}_\omega$ are related to the Minkowski operator by
\begin{align}\label{eqn:MinkowskiUnruh}
    \hat{a}_{kl} &= \int \D \omega \Big( A_{k\omega} \hat{c}_{\omega l} + \hat{B}_{k\omega} \hat{d}_{\omega l} \Big) ,
    \\
    \hat{a}_{kr} &= \int \D \omega \Big( B_{k\omega} \hat{c}_{\omega r} + \hat{A}_{k\omega} \hat{d}_{\omega r} \Big) ,
\end{align}
where
\begin{align}
    A_{k\omega} = B_{k\omega}^\star &= \frac{ i \sqrt{\sinh (\pi \omega/a)}}{2\pi \sqrt{\omega k }} \Gamma( 1- i \omega/a) \left( \frac{k}{a}\right)^{i\omega/a} 
\end{align}
are the Bogoliubov transformation coefficients \cite{crispino2008unruh}. The Unruh operators are a stepping stone between the Rindler and Minkowski operators. 

\subsection{Rindler Wedges, Coordinates and Operators}
\begin{figure}[h]
    \centering
    \includegraphics[width=0.9\linewidth]{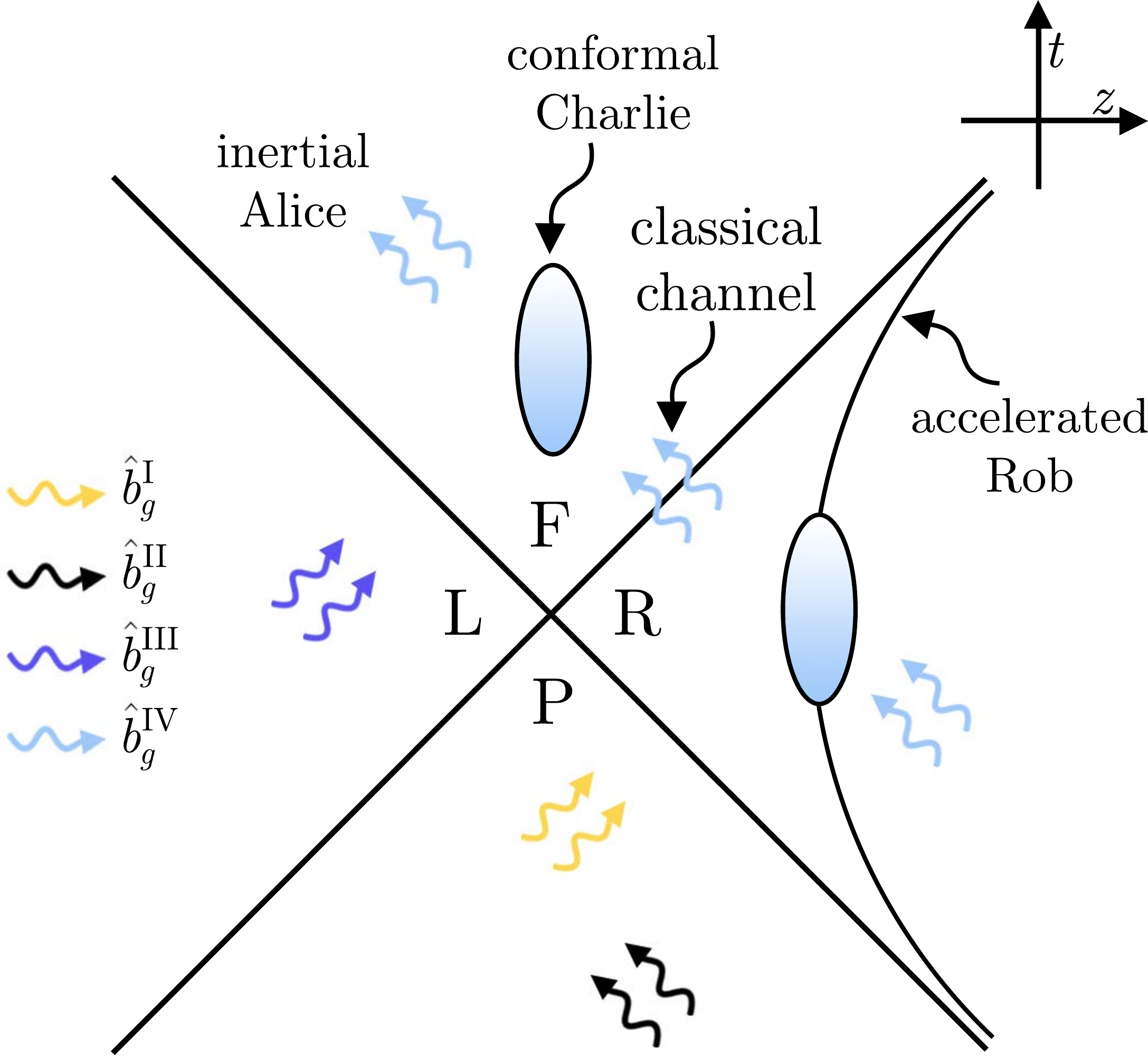}
    \caption{Rob, accelerating uniformly in the right Rindler wedge, interacts with the Rindler wavepacket modes $\hat{b}_g^\mathrm{IV}$ and $\hat{b}_g^\mathrm{I}$ in the blue ellipse region. The classical channel is sent to the conformal receiver Charlie, who mixes it with the right-moving Rindler modes $\hat{b}_g^\mathrm{III}$ at a beam splitter (graded ellipse), before transmitting the state to Alice, in the Minkowski vacuum. The unitary evolution of the Rindler modes is shown in Fig.\ \ref{fig:circ}.}
    \label{fig:2}
\end{figure}

Minkowski spacetime can be partitioned into four wedges denoted (R), (L), (F) and (P), representing the right and left Rindler wedges and the future and past light cones respectively (Fig.\ \ref{fig:2}) \cite{su2017quantum}. 
coordinates in these wedges are defined as follows,
\begin{align}\label{eqn:global1}
\begin{split}
\mathrm{(F)}&\:\:\:\begin{cases}
t = a^{-1} e^{a\eta}\cosh (a\zeta),
\vt 
\\
z = a^{-1} e^{a\eta}\sinh (a\zeta) ,
\vt 
\end{cases}
\end{split} 
\\
\begin{split}
\mathrm{(P)}&\:\:\:
\begin{cases}
t = -a^{-1} e^{a\bar{\eta}} \cosh (a\bar{\zeta}),
\vt 
\\
z = - a^{-1} e^{a\bar{\eta}} \sinh (a\bar{\zeta}), 
\vt 
\end{cases}
\end{split} 
\end{align}
where $\eta$, $\bar{\eta}$ and $\zeta, \bar{\zeta}$ are the conformal time and spatial coordinates of observers restricted to the $(\mathrm{F})$ and $(\mathrm{P})$ light-cones respectively, and
\begin{align}\label{eqn:global2}
\begin{split}
\mathrm{(R)}& \:\:\:
\begin{cases}
t = a^{-1} e^{a\xi} \sinh (a\tau) ,
\vt
\\
z = a^{-1} e^{a\xi} \cosh (a\tau) ,
\vt 
\end{cases}
\end{split}
\\
\begin{split}
\mathrm{(L)}&\:\:\:
\begin{cases}
t = - a^{-1} e^{a\bar{\xi}} \sinh (a\bar{\tau}),
\vt 
\\
z = -a^{-1} e^{a\bar{\xi}} \cosh (a\bar{\tau}),
 \vt 
\end{cases}
\end{split},
\end{align}
where $\tau,\bar{\tau}$ and $\xi,\bar{\xi}$ are the proper time and the spatial coordinates of a uniformly accelerated observer with proper acceleration $a$ in the $(\mathrm{R})$ and $(\mathrm{L})$ wedge respectively. In Fig.\ \ref{fig:2}, the conformal receiver Charlie is stationary according to an inertial observer but operates a beam splitter with time-dependent reflectivity. Specifically, Olson et al. demonstrated in \cite{olson2011entanglement} that an Unruh-deWitt detector whose energy gap is scaled continuously as $1/at$ responds to the Minkowski vacuum identically to one on an accelerated trajectory with fixed energy gap. By using an analogously time-dependent, mode-selective beam splitter, Charlie remains in the causal future of Rob, who is uniformly accelerating, but can still interact with the same modes as him. This proceeds by analytically extending the (left-moving) Rindler mode functions in $(\mathrm{R})$ into $(\mathrm{P})$. 

To see this, it is useful to introduce the Rindler analogues to the light-cone coordinates, given by 
\begin{align}
    \begin{split}
        \mathrm{(F)} \:\:\nu &= \eta + \zeta 
        \vt ,
        \\
        \mathrm{(P)} \:\:\overline{\nu} &= - \overline{\eta} - \overline{\zeta} 
        \vt , 
        \\
        \mathrm{(R)} \:\: \chi &= \tau + \xi 
        \vt, 
        \\
        \mathrm{(L)} \:\:\overline{\chi} &= - \overline{\tau} - \overline{\xi} 
        \vt , 
    \end{split}
        \begin{split}
       \mu &= \eta - \zeta 
       \vt, 
       \\
       \overline{\mu} &= - \overline{\eta} + \overline{\zeta} 
       \vt , 
       \\
       \kappa &= \tau - \xi \vt , 
       \\
       \overline{\kappa} &= - \overline{\tau} + \overline{\xi} .
       \vt 
    \end{split}
\end{align}
In the reference frame of a uniformly accelerated observer, the field operator can be expanded in Rindler modes confined to the $(\mathrm{R})$ and $(\mathrm{L})$ wedges of spacetime (considering for simplicity the left-moving sector of the field, since in (1+1)-dimensions this decouples from the right-moving sector):
\begin{align}
    \hat{\Phi}(V) &=  \int \D \omega \Big(  \hat{b}_{\omega l}^\mathrm{R} g_{\omega l}^\mathrm{R} + \hat{b}_{\omega l}^L g_{\omega l}^\mathrm{L} + \mathrm{h.c} \Big) 
\end{align}
where $\omega$ is the frequency defined with respect to the proper time co-ordinate in the respective quadrant of Rindler space. Equivalently, $\hat{\Phi}$ can be expanded in the modes restricted to the future $(\mathrm{F})$ and past $(\mathrm{P})$. For example, the left-moving mode functions are 
\begin{align}
    g^\mathrm{R}_{\omega l}(\chi) &= \frac{1}{\sqrt{4\pi\omega}} e^{-i\omega \chi} , 
\vt 
\\ 
    g_{\omega l}^\mathrm{L} (\bar{\chi}) &= \frac{1}{\sqrt{4\pi\omega}} e^{-i\omega\bar{\chi}} 
    \vt , 
\end{align}
with their analogues in the future and past given by,
\begin{align}
    g_{\omega l}^\mathrm{F}(\nu) &= \frac{1}{\sqrt{4\pi\omega}} e^{-i \omega \nu}
    \vt, 
    \\
    g_{\omega l}^\mathrm{P}(\overline{\nu})&= \frac{1}{\sqrt{4\pi\omega}} e^{-i \omega \overline{\nu}}.
    \vt 
\end{align}
It can be easily shown that
\begin{align}
    g_{\omega l}^\mathrm{R}(V) = g_{\omega l}^\mathrm{F}(V) &= \frac{1}{\sqrt{4\pi\omega}} (aV)^{-i\omega/a}\Theta(V) \vt 
    , 
    \\
    g_{\omega l}^\mathrm{L}(V) = g_{\omega l}^\mathrm{P} (V)  &= \frac{1}{\sqrt{4\pi\omega}} (-aV)^{i\omega/a}\Theta(-V).
    \vt 
\end{align}
Because $g_{\omega l}^\mathrm{R}(\chi)$ and $g_{\omega l}^\mathrm{F}(\nu)$ are identical as functions of $V$, as are $g_{\omega l}^\mathrm{L}(\chi)$ and $g_{\omega l}^\mathrm{P}(\nu)$, the Bogoliubov coefficients relating the Minkowski and Rindler modes in $(\mathrm{R})$ and $(\mathrm{L})$ are duplicated in $(\mathrm{F})$ and $(\mathrm{P})$. Therefore the Rindler mode functions can be extended across the Rindler horizons by a change of coordinates. Using this property, one can define left-moving mode functions which span the entirety of Regions $\mathrm{II}$ and $\mathrm{IV}$ in Minkowski spacetime \cite{higuchi2017entanglement,olson2011entanglement}, with their corresponding bosonic operators given by
\begin{align} \label{eq:global1}
    \hat{b}_\omega^\mathrm{II} &:= \hat{b}_{\omega l
    }^\mathrm{P} = \hat{b}_{\omega l}^\mathrm{L} ,
    \vt 
    \\ \label{eq:global2}
    \hat{b}_\omega^\mathrm{IV} &:= \hat{b}_{\omega l}^\mathrm{R} = \hat{b}_{\omega l}^\mathrm{F} ,
    \vt 
\end{align}
which have support over the entirety of Regions II and IV (see Fig \ref{fig:2}). For completeness, the following association for the right-moving Rindler operators can also be made,
\begin{align}\label{34}
    \hat{b}_\omega^\mathrm{I} &:=  \hat{b}_{\omega r}^\mathrm{L} = \hat{b}_{\omega r}^\mathrm{F} ,
    \vt 
    \\
    \hat{b}_\omega^\mathrm{III} &:= \hat{b}_{\omega r}^\mathrm{P} = \hat{b}_{\omega r}^\mathrm{R} . 
    \vt 
    \label{35}
\end{align}
Thus, the left-moving sector of the field can be expanded as follows \cite{higuchi2017entanglement}
\begin{align}
    \hat{\Phi} (V) &= \sum_\mathrm{\chi = II,IV} \int_0^\infty \D \omega \Big( \hat{b}_{\omega l}^\chi g_{\omega l}^\chi (V) + \mathrm{h.c} \Big) . 
    \vt 
\end{align}
\noindent Demonstrating the presence of entanglement between the left-moving modes in $\mathrm{II}$ and $\mathrm{IV}$ (analogously the right-moving modes in $\mathrm{I}$ and $\mathrm{III}$) is identical to the derivation for spacelike entanglement between $(\mathrm{R})$ and $(\mathrm{L})$ with a mere change of labels: that is, $(\mathrm{P})$, $(\mathrm{L})$ $\Rightarrow$ $\mathrm{II}$ and $(\mathrm{R})$, $(\mathrm{F})$ $\Rightarrow$ $\mathrm{IV}$ (for the right-movers, replacing $(\mathrm{P})$, $(\mathrm{R})$ $\Rightarrow$ I and $(\mathrm{L})$, $(\mathrm{F})$ $\Rightarrow$ $\mathrm{III}$) \cite{olson2011entanglement,higuchi2017entanglement,crispino2008unruh}. 
\noindent This derivation is shown in detail in \cite{olson2011entanglement}. Eq.\ (\ref{eq:global1})-(\ref{35}) imply that the complete sets of right-moving modes in Regions $\mathrm{I}$ and $\mathrm{III}$ and the left-moving modes in Regions II and IV are entangled, in the same way that spacelike separated Rindler modes in $(\mathrm{R})$ and $(\mathrm{L})$ and timelike separated conformal modes in $(\mathrm{F})$ and $(\mathrm{P})$ are entangled. In the following analysis, we harness this entanglement as the resource for teleportation between the accelerated and stationary observers. 

\section{Quantum Circuit Model}\label{sec:IV}
The quantum circuit model is a nonperturbative approach to describing interactions between observers in relativistic, non-inertial reference frames and quantum fields \cite{su2017quantum,su2019decoherence}. It describes these interactions as the Heisenberg evolution of Unruh operators ($\hat{c}_{\omega i},\hat{d}_{\omega i}$) into Rindler operators $(\hat{b}_{\omega}^\chi$) with $\chi = \mathrm{I}, \mathrm{II}, \mathrm{III} , \mathrm{IV}$, and back into Unruh operators, which can then be detected as Minkowski modes according to Eq.\ (\ref{eqn:MinkowskiUnruh}). The basis transformation between the Unruh and Rindler operators is defined in \cite{su2017quantum} and is essentially two-mode squeezing, the `source' of entanglement. 

\begin{figure}[h]
    \centering
    \includegraphics[width=0.9\linewidth]{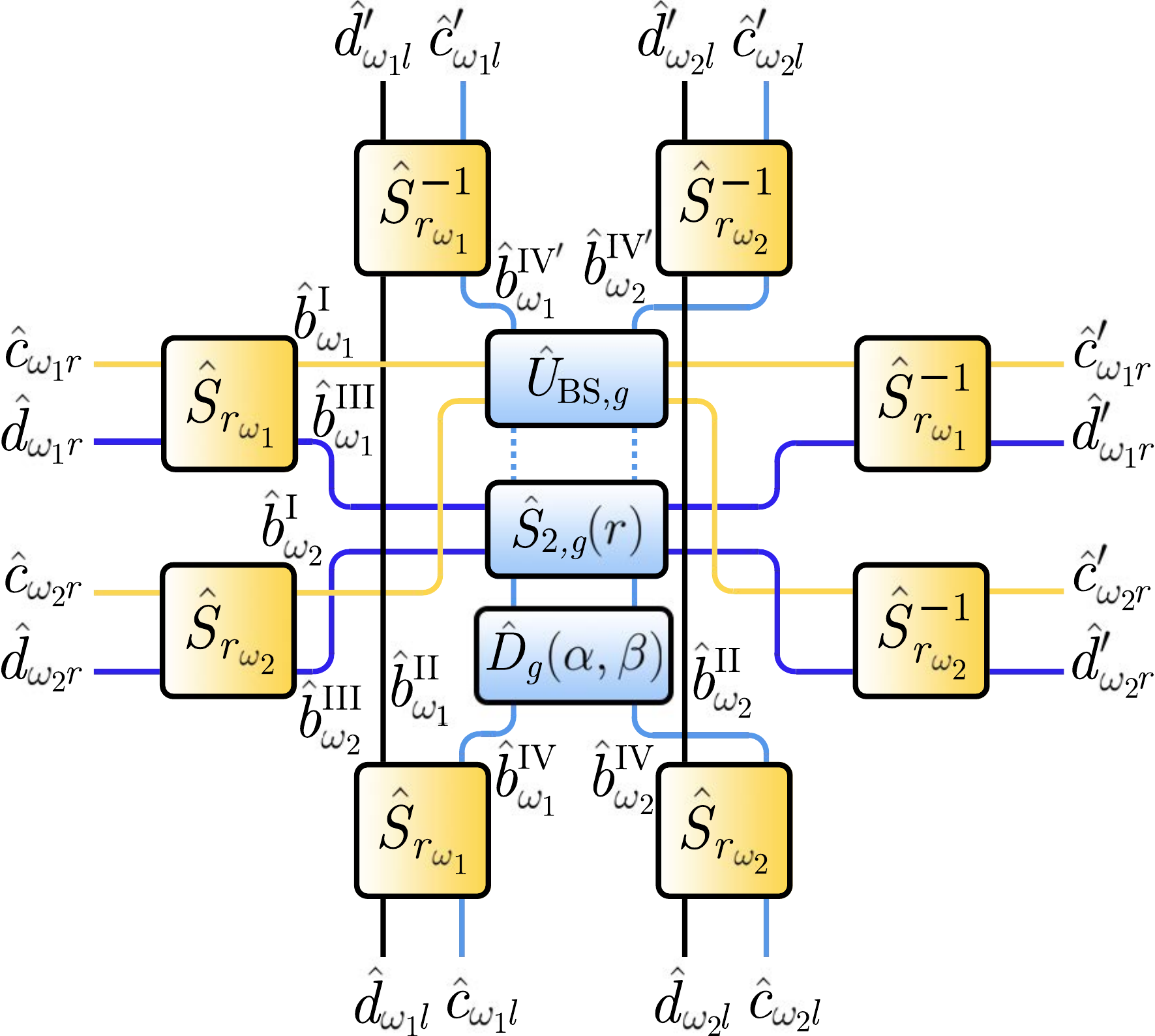}
    \caption{Quantum circuit for teleportation between the accelerated sender and conformal receiver. The time-dependent interaction mixes different single-frequency Rindler modes, and the coloured lines identify Rindler modes confined to different regions of spacetime (these match Fig.\ \ref{fig:2}). The accelerated observer creates the signal using $\hat{D}_g(\alpha,\beta)$, amplifies this through $\hat{S}_{2,g}(r)$, before the conformal receiver attenuates it using a beam splitter, $\hat{U}_\mathrm{BS,g}$. The classical channel is denoted by the dashed line.}
    \label{fig:circ}
\end{figure}

The circuit in Fig.\ \ref{fig:circ} imposes the all-optical teleportation protocol into the reference frames of accelerated Rob and conformal Charlie. We consider interactions with localised wavepacket modes in the accelerated (and conformal) frames, described by the operator
\begin{align}
    \hat{b}_{g}^\chi &\equiv \int_0^\infty \D \omega\: g(\omega) \hat{b}_{\omega }^\chi,
\end{align}
where $g(\omega)$ assumes a Gaussian profile. The circuit diagram illustrates transformations of many single-frequency modes and the corresponding interaction in the accelerated frame as acting on a single, localised wavepacket mode, in the continuum limit.  
For the unitary interaction $\hat{U}_g$ with an arbitrary Rindler wavepacket $g(\omega)$, there exist a complete set of wavepackets $g_{\perp,i}(\omega)$ orthogonal to $g(\omega)$ which do not interact with $\hat{U}_g$. Thus, the single-frequency Rindler operators can be decomposed into an `interacting' and `noninteracting' part \cite{rohde2007spectral},
\begin{align}\label{sumeqn}
    \hat{b}_\omega^\chi &= g^\star(\omega) \hat{b}_g^\chi + \sum_i g_{\perp,i}(\omega) \hat{b}_{g\perp,i}^\chi.
    \vt 
\end{align}
Acting $\hat{U}_g$ on a single-frequency Rindler operator yields
\begin{align}\label{eqn:Rindlerwavepacketeqn}
    \hat{b}_\omega^{\chi\prime} = \hat{U}_g^\dd \hat{b}_\omega^\chi \hat{U}_g &= \hat{b}_\omega^\chi + g^\star(\omega) \Big( \hat{U}_g^\dd \hat{b}_g^\chi \hat{U}_g - \hat{b}_g^\chi \Big) ,
    \vt 
\end{align}
from Eq.\ (\ref{sumeqn}). Using the relations between the single-frequency Unruh and Rindler operators \cite{su2017quantum} and Eq.\ (\ref{sumeqn}) and (\ref{eqn:Rindlerwavepacketeqn}), we obtain the following expressions for the output left-moving Unruh operators \cite{su2019decoherence}
\begin{align} \label{eqn:unruh1}
    \hat{c}_{\omega l}'&= \hat{c}_{\omega l} + g^\star(\omega) \cosh r_\omega \big( \hat{U}_g^\dd \hat{b}_g^\mathrm{IV} \hat{U}_g - \hat{b}_g^\mathrm{IV} \big) , 
    \vt 
    \\ \label{eqn:unruh2}
    \hat{d}_{\omega l}' &= \hat{d}_{\omega l} - g(\omega) \sinh r_\omega  \big( \hat{U}_g^\dd \hat{b}_g^\mathrm{IV\dd} \hat{U}_g - \hat{b}_g^\mathrm{IV\dd} \big) . 
    \vt 
\end{align}
where $ r_\omega = \mathrm{arctanh}\: e^{-\pi\omega/a}$ is the two-mode squeezing factor. Alice reconstructs the Minkowski modes from $\hat{c}_{\omega l}', \hat{d}_{\omega l}'$. In Fig.\ \ref{fig:circ}, $\hat{D}_g(\alpha,\beta) = \hat{D}_g(\beta) \hat{D}_g(\alpha)$ where $\hat{D}_g(\alpha)$ imposes a strong local oscillator to the classical signal $\beta$, prepared and teleported by Rob, which is an important feature of the self-homodyne detection model which we employ. 

\subsection{Self-Homodyne Detection}
Since the teleported state is unknown to the inertial observer, we model them as operating an infinite bandwidth detector that integrates over \textit{all} Minkowski modes,  
\begin{align}\label{particles}
    \hat{N} &= \int_{-\infty}^\infty  \mathrm{d} k \:\hat{a}_k^\dd \hat{a}_k.
\end{align}
This approximates a physical detector that captures a finite range of frequencies much larger than the spread of frequencies of the signal \cite{onoe2018particle,su2019decoherence}. By displacing the input Rindler mode by a strong local oscillator characterised by the complex amplitude $\alpha = |\alpha| e^{i\phi}$, the quadrature amplitudes of the input field now exhibit  phase-dependent oscillations,
\begin{align}
    \hat{X}(\phi) &= \hat{a} e^{-i\phi} + \hat{a}^\dd e^{i\phi},
    \vt 
\end{align}
where $\phi$ is the phase of the displacement in phase space. In the limit where $|\alpha|\gg  1$, $\hat{N}(\phi)$ can be approximated as
\begin{align}
    \hat{N}(\phi) &\simeq \al + |\alpha|(\beta + \beta^\star).
    \vt 
\end{align}
The quadrature operator can be written as
\begin{align}
    \langle \hat{X}(\phi) \rangle &\simeq \frac{\langle \hat{N} ( \phi) \rangle - \langle \hat{N}_0\rangle }{\sqrt{\langle \hat{N}_0 \rangle }},
    \vt 
\end{align}
where $\langle\hat{N}_0\rangle$ is the average number of photons when no signal $\beta$, is imposed. The variance of the quadrature amplitude is  \cite{onoe2018particle,su2019decoherence}
\begin{align}\label{60}
    \left( \Delta X(\phi) \right)^2 &= \frac{\langle \hat{N}^2(\phi) \rangle - \langle \hat{N} (\phi) \rangle^2}{\langle \hat{N}_0 \rangle} = \frac{\left( \Delta N(\phi) \right)^2}{\langle \hat{N}_0 \rangle},
    \vt 
\end{align}
where $\Delta N ( \phi)$ is the variance in the photon number according to the inertial detector. Conveniently, the purity of Gaussian states is quantified by Eq.\ (\ref{60}), the criterion for which is that the product of the $\phi = 0$ and $\phi = \pi/2$ quadrature variances is equal to unity \cite{bachor2004guide}. 

\section{Analytic Results}\label{sec:V}
In the following analysis, we consider the teleportation of the following states from the accelerated frame:
\begin{enumerate}[label=(\alph*)]
    \item  a displaced Minkowski vacuum state (a displaced thermal state in the Rindler observer's frame, acting as a classical signal) created by displacing $\smash{\hat{b}_{g}^\mathrm{IV}}$ with $\smash{\hat{D}_g(\beta)}$ (Fig.\ \ref{fig:circ}), and
    \item a squeezed thermal state (a quantum signal) generated by single-mode squeezing the input Rindler mode $\hat{b}_g^\mathrm{IV}$,
\end{enumerate}
and determine their purity according to Alice. For (a), an ideal teleportation protocol should result in Alice detecting a pure coherent state. We expect this from the result in \cite{su2019decoherence}, which found that \textit{direct transmission} of the Rindler-displaced Minkowski vacuum was detected as a pure coherent state according to an inertial Minkowski observer. 

\subsection{Displaced Thermal State}
From Fig.\ \ref{fig:circ}, $\hat{b}_g^\mathrm{IV}$ evolves under the action of the unitary $\smash{\hat{U}_g = \hat{U}_\mathrm{BS,g} \hat{S}_{2,g} \hat{D}_g(\alpha,\beta)}$ as 
\begin{align}\label{eqn:heisenbergevolution}
   \hat{b}_g^\mathrm{IV\prime}
    &= \hat{b}_g^\mathrm{IV} + \hat{b}_g^\mathrm{III\dd} \tanh (r)
    \non 
    \\
    & \qquad - \hat{b}_g^\mathrm{I} \sqrt{1 - \frac{1}{\cosh^2(r)}} + \alpha + \beta ,
\end{align}
where $\alpha$ characterises the strong local oscillator, and $\beta$ (with $|\beta|\ll |\alpha|$) creates the displaced state which we analyse. Using Eq.\ (\ref{eqn:Rindlerwavepacketeqn}) and assuming the limit of strong amplification of the field through the classical channel $(r\gg 1$) the output single-frequency Rindler operator takes the form
\begin{align} \label{eqn:57}
\hat{b}_\omega^\mathrm{IV\prime}     &\simeq  \hat{b}_\omega^\mathrm{IV} + g^\star(\omega) \left( \hat{b}_g^\mathrm{III\dd}  - \hat{b}_g^\mathrm{I} + \alpha + \beta \right) .
\end{align}
Since the left-moving modes originating in the past light cone do not interact with $\hat{U}_g$, then $\hat{b}_g^\mathrm{II\prime} = \hat{b}_g^\mathrm{II}$. From Eq.\ (\ref{eqn:57}) we find that Charlie, in the conformal reference frame, sees an approximation of the original state, which becomes better as the acceleration increases. This becomes evident after decomposing the Rindler operators into their constituent Unruh operators and assuming $g^\star(\omega) = g(\omega)$, such that,
\begin{align}\label{eqn:conformal}
    \hat{b}_\omega^\mathrm{IV} {}' &\simeq \hat{b}_{\omega}^\mathrm{IV} +  g(\omega) \Big( \varphi_\mathrm{CS}  (\hat{d}_{\omega' r}^{\dd} - \hat{c}_{\omega' r} )  + \alpha  + \beta \Big) ,
    \vt 
\end{align}
where
\begin{align}
   \varphi_\mathrm{CS} &= \int\D \omega \:g(\omega) \big( \cosh (r_{\omega}) - \sinh (r_{\omega}) \big) ,
\end{align} 
which vanishes in the limit of infinite acceleration. Eq.\ (\ref{eqn:conformal}) is formally equivalent to Eq.\ (\ref{eqn:9}) for the static case. In the limit of infinite acceleration, the right-moving Rindler modes become perfectly entangled, reducing the output Rindler operator to an identical reconstruction of the input displaced by the amplitude $g(\omega)(\alpha + \beta)$. Therefore in the limit of infinite acceleration, our model predicts the ideal teleportation of states sent between the uniformly accelerated sender in the right Rindler wedge and the conformal receiver restricted to the future. We note however, that for large accelerations, the state becomes increasingly thermalised. 

\begin{figure}[h]
    \centering
    \includegraphics[width=\linewidth]{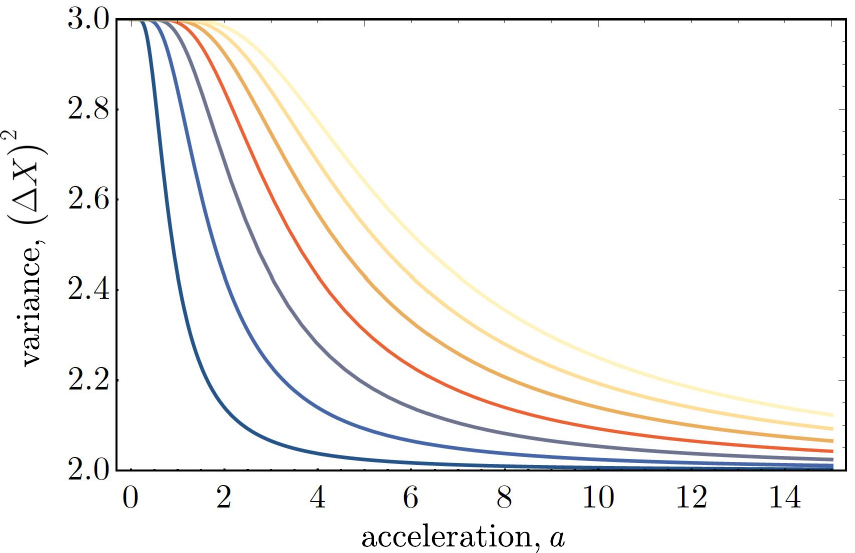}
    \caption{$(\Delta X)^2$ for the teleported state as detected by Alice in the Minkowski vacuum, as a function of the acceleration. The individual lines represent $\omega_0 = 0.5, 1.0, \hdots, 3.5$ (increasing from blue to yellow). }
    \label{fig:variance1}
\end{figure}

We now consider the purity of the output state according to an inertial observer with access to the entire Minkowski spacetime, which requires a transformation of the Rindler operators into Minkowski operators. Applying Eq.\ (\ref{eqn:unruh1}) and (\ref{eqn:unruh2}) to the output Rindler operators yields
\begin{align}     
    \hat{c}_{\omega l}' &= \hat{c}_{\omega l} + g(\omega) \cosh (r_\omega) \big( \hat{b}_g^\mathrm{III\dd} - \hat{b}_g^\mathrm{I} + \alpha + \beta \big) 
    \vt ,
\\
     \hat{d}_{\omega l}'&= \hat{d}_{\omega l} - g(\omega) \sinh (r_\omega) \big( \hat{b}_g^\mathrm{III} - \hat{b}_g^\mathrm{I\dd} + \alpha^\star + \beta^\star  \big) . 
     \vt 
\end{align}
Eq.\ (\ref{particles}) for the Minkowski photon number operator simplifies to
\begin{align}
    \hat{N} &= \int \D \omega \big( \hat{c}_{\omega l}^{\dd\prime} \hat{c}_{\omega l}' + \hat{d}_{\omega l}^{\dd\prime} \hat{d}_{\omega l}' \big) ,
\end{align}
where we have used the identities $\int \D k A_{k\omega}A_{k\omega'}^\star = \delta( \omega - \omega')$ and $\int \D k \: A_{k\omega}A_{k\omega'}= 0$. Similarly, the square of the number operator is
\begin{align}
\hat{N}^2 &= \int \D \omega \int  \D \gamma  \big(c_{\omega l}^{\dd\prime} c_{\omega l}' + d_{\omega l}^{\dd\prime} d_{\omega l}' \big) \big( c_{\gamma l}^{\dd\prime} c_{\gamma l}' + d_{\gamma l}^{\dd\prime} d_{\gamma l}'\big) .
\end{align}
By computing the relevant vacuum expectation values of the products of output Unruh operators, the quadrature variance of the output state detected by Alice is found to be (details in Appendix A) \begin{align}\label{eqn:allopticalquadvar}
    \left( \Delta X \right)^2 &= \underbrace{ 2 \mathcal{I}_\mathrm{CS} \big( \mathcal{I}_\mathrm{C} + \mathcal{I}_\mathrm{S} \big)}_\mathrm{thermal \: noise}\: + \underbrace{1\vphantom{\big|}}_\mathrm{QNL} 
\end{align}
where
\begin{align}
    \mathcal{I}_\mathrm{C} &= \int\D \omega\:g(\omega)^2 \cosh^2 (r_\omega)
    \vt, 
\\
    \mathcal{I}_\mathrm{S} &= \int\D\omega\:g(\omega)^2 \sinh^2 (r_\omega),
\vt 
\\
    \mathcal{I}_\mathrm{CS} &= \int\D\omega\: g(\omega)^2 \big( \cosh (r_\omega) - \sinh (r_\omega) \big)^2 .
\vt 
\end{align}
In the limiting case where $g(\omega)$ is a narrow bandwidth Rindler mode $(\sigma \ll \omega_0)$, the quadrature variance reduces to 
\begin{align}\label{eqn:coherentquadvar}
    \left(\Delta X \right)^2 &\simeq  \big( 1 + \exp ( - 4r_0  ) \big) + 1 .
    \vt 
\end{align}
For $\omega_0/a\gg 1$, the non-inertial (Charlie and Rob) observers are effectively reduced to a static, inertial frame. As shown in Fig.\ \ref{fig:variance1}, this occurs in the limits $a\to 0$ with fixed $\omega_0$ (smaller accelerations) or $\omega_0\gg 1$ with fixed $a$. In the latter case, the non-inertial observers interact with sufficiently high-frequency modes such that they are surrounded by vacuum (due to the exponentially decaying tail of the Planck spectrum). The input Rindler modes remain independent, so that two additional QNL are added to the output state, analogous to Eq.\ (\ref{eqn:6}).  
\noindent As $\omega_0/a\to0$, the right-moving resource modes become perfectly entangled. Curiously, the output state detected by the inertial Minkowski observer is not pure, but carries at best, an extra unit of QNL above the Heisenberg limit of unity. Even for a perfect entanglement resource, the state appears mixed according to the inertial observer, despite a reduction below the classical limit.

To understand this behaviour, we notice that Eq.\ (\ref{eqn:allopticalquadvar}) contains a thermal noise term in addition to the single QNL that originates from the Rindler mode displaced by Rob and teleported to Charlie. It was found in \cite{su2019decoherence} that the direct transmission of a displaced Rindler mode from the accelerated reference frame was detected as a pure coherent state according to an inertial Minkowski observer. Comparatively, the thermal noise term in Eq.\ (\ref{eqn:allopticalquadvar}) approaches one (an additional QNL) rather than vanishing in the limit of infinite acceleration, since the output Rindler mode \textit{only reconstructs the input identically at infinite acceleration}. Hence, whilst the noise terms from the right-moving vacuum modes become \textit{increasingly suppressed} with increasing acceleration (due to stronger entanglement between the resource modes $\hat{b}_g^\mathrm{I}$, $\hat{b}_g^\mathrm{III}$), they are \textit{simultaneously amplified} due to the increasing thermalisation of the Unruh effect. These competing effects fortuitously balance one another, resulting in (for the ideal case of infinite acceleration) an additional QNL to the output state. 

\subsection{Squeezed Thermal State}
To examine the evolution of nonclassical states through the teleportation protocol, we apply single-mode squeezing to the initial state prepared in the accelerated reference frame. We are also interested in how the teleported state is affected by the so-called decoherence effects found previously in \cite{su2019decoherence,onoe2018particle}. As before, we derive the output single-frequency Rindler operators, given by 
\begin{align}
    \hat{b}_\omega^\mathrm{IV} {}' &= b_\omega^\mathrm{IV} + g(\omega) \left( b_g^\mathrm{IV} (\cosh r - 1) \right. 
    \vt 
\nonumber \\
    & \qquad \left. + b_g^\mathrm{IV\dd} \sinh r + b_g^\mathrm{I\dd} - b_g^\mathrm{III} + \alpha \right) ,
\vt 
\end{align}
where now $\hat{U}_g = \hat{U}_\mathrm{BS} \hat{S}_{2,g}(r) \hat{S}_{1,g}(r) \hat{D}_g(\alpha)$ and $\hat{S}_{1,g}(r)$ squeezes the Rindler vacuum. It is straightforward to show that the output Unruh operators are
\begin{align}    
    \hat{c}_{\omega l}' &= \hat{c}_{\omega l} + g(\omega) \cosh r_\omega \big( \hat{b}_g^\mathrm{IV} (\cosh r - 1) \nonumber 
    \vt 
\\
    &  + \hat{b}_g^\mathrm{IV\dd} \sinh r + \hat{b}_g^\mathrm{III\dd} - \hat{b}_g^\mathrm{I} + \alpha \big),
\vt 
\\ 
    \hat{d}_{\omega l}' &= \hat{d}_{\omega l} - g(\omega) \sinh r_\omega \big( \hat{b}_g^\mathrm{IV\dd} (\cosh r - 1)  \nonumber 
\vt
\\
    &  + \hat{b}_g^\mathrm{IV} \sinh r + b_g^\mathrm{III} - \hat{b}_g^\mathrm{I\dd} + \alpha^\star \big).  
\vt 
\end{align}
Performing a similar calculation to the previous case, we find that the quadrature variance of the state detected by the inertial Minkowski observer is given by,
\begin{align}\label{eqn:squeezingvar}
    \left( \Delta X (\phi) \right)^2 &= \underbrace{2\mathcal{I}_\mathrm{CS} \big( \mathcal{I}_\mathrm{C} + \mathcal{I}_\mathrm{S} \big)}_\mathrm{thermal \: noise} + \underbrace{\Delta(\phi) \vphantom{\big|}}_\mathrm{decoherence} ,
\end{align}
where 
\begin{align}\label{75}
    & \Delta(\phi) =
    \non 
    \vt 
\\
    & \:\:\: 
    \cosh (2r) + 4\mathcal{I}_\mathrm{C} ( \mathcal{I}_\mathrm{C} - 1) ( \cosh (2r) - 2 \cosh (r) + 1 ) \nonumber 
\vt 
\\
    & \:\:\: + 2 \sinh (r) \Big[( 2 \mathcal{I}_\mathrm{C} - 1)^2 \cosh (r) - 4\mathcal{I}_\mathrm{C} ( \mathcal{I}_\mathrm{C} - 1) \Big] \cos(2\phi) .
\vt 
\end{align}
The maximum and minimum quadrature variances correspond to the phases $\phi = 0, \pi/2$,
\begin{align} \label{eqn:Variance1}
    \left( \Delta X (0) \right)^2 &= 2\mathcal{I}_\mathrm{CS} (\mathcal{I}_\mathrm{C} + \mathcal{I}_\mathrm{S}) + \Delta(0) 
    \vt ,
    \\ \label{eqn:Variance2}
    \left( \Delta X (\pi/2) \right)^2 &=  2\mathcal{I}_\mathrm{CS} (\mathcal{I}_\mathrm{C} + \mathcal{I}_\mathrm{S}) + \Delta(\pi/2)
    \vt ,
\end{align}
where
\begin{align}\label{eqn:70}
   \Delta(0) &=  e^{2r} + 4 \mathcal{I}_\mathrm{C}( \mathcal{I}_\mathrm{C} -1) (e^{r} - 1)^2  
   \vt ,
   \\ \label{eqn:71}
   \Delta(\pi/2) &= e^{-2r} + 4 \mathcal{I}_\mathrm{C} (\mathcal{I}_\mathrm{C} -1 ) (e^{-r}-1)^2.
   \vt 
\end{align}
Eq.\ (\ref{eqn:squeezingvar}) reveals two competing factors affecting the purity of the output state. The first is the thermal noise term found previously for the displaced Rindler vacuum state prepared and teleported by Rob, which is uncorrelated from the noise of the signal itself. The second is the decoherence term $\Delta(\phi)$ which also appeared in \cite{su2019decoherence} for a squeezed state sent from the accelerated frame to an inertial observer. There, it was concluded that such terms arise from the transformation of the bipartite Hilbert space of the Rindler and Unruh modes to the single Hilbert space of the Minkowski modes. This leads to a loss of phase information in the Unruh modes when computing the Minkowski particle number and the observed decoherence according to inertial detectors. 
\begin{figure}[h]
    \centering
    \includegraphics[width=\linewidth]{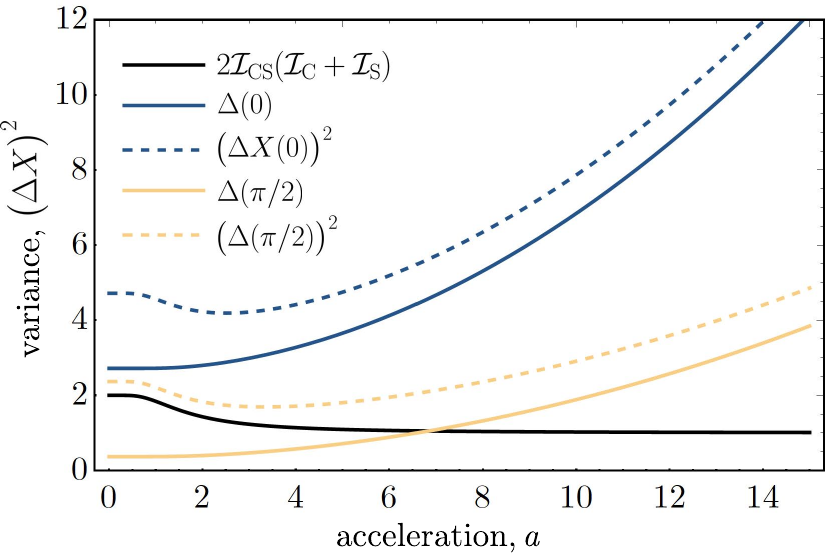}
    \caption[Quadrature variances of the output state after all-optical teleportation of a quantum signal]{Contributions to $(\Delta X)^2$ for the squeezed thermal state as detected by the inertial observer in the Minkowski vacuum, with $r= 0.5, \omega_0 = 1$. The decoherence terms dominate for large $a$. }
    \label{fig:Variance2}
\end{figure}
Fig.\ \ref{fig:Variance2} shows that for small $a$, the strengthening of entanglement between the Rindler modes reduces the noise on the output state, but as $a$ increases, the Unruh effect becomes significant and the variances grow unbounded. We conclude that the purity of the output state according to Alice, in the Minkowski vacuum, is affected by
\begin{enumerate}
\item the thermal noise term, which encodes within it the competing suppression and amplification effects discussed previously, and 
\item the squeezing of the input Rindler mode, the source of the decoherence terms first found in \cite{su2019decoherence}.
\end{enumerate}

\section{Conclusion}\label{sec:VI}
The decoherence of quantum states prepared in an accelerated reference frame and detected by inertial observers has been studied previously \cite{su2019decoherence,onoe2018particle}. In \cite{su2019decoherence}, the authors argue that any interaction that leads to entanglement between the Unruh modes, as occurs for the squeezed Rindler vacuum state, appears as decoherence according to the measurements of inertial observers, whilst classical signals do not produce such entanglement. Nevertheless, we found that the displaced Rindler vacuum state prepared and \textit{teleported} by Rob was still polluted by additional thermal noise terms according to the inertial Minkowski observer. We attributed this additional decoherence to the fact that the output state teleported to Charlie only becomes an identical reconstruction of the input at infinite acceleration, which is nullified by the opposing thermalisation effects of Unruh radiation in this limit. 

As is evident from the simplicity of our calculations, the quantum circuit model presents a powerful tool for analysing quantum information protocols in relativistic, non-inertial settings, possessing some advantages over previous models. A prominent one is that the model is nonperturbative, which allows the evolution of quantum fields to be described by unitary operations. It is also a natural setting to analyse continuous-variable protocols where only discrete-variable versions have been studied, and these can be straightforwardly mapped to quantum optical contexts. An immediate extension to this work would be to implement other entanglement-based protocols using the quantum circuit model, such as continuous-variable dense coding \cite{braunstein2000dense,ralph2002unconditional}, quantum energy teleportation \cite{hotta2009quantum,hotta2010controlled,hotta2014quantum,funai2017engineering,verdon2016asymptotically} and quantum key distribution \cite{ralph2015quantum}. 

\section{Acknowledgements}
This research is supported by the Australian Research Council (ARC) under the Centre of Excellence for Quantum Computation and Communication Technology (Grant No. CE170100012).

\begin{widetext}
\appendix

\section{Rindler-Displaced Minkowski Vacuum State}
\noindent In the main text, we derived expressions for the output Unruh operators, given by 
\begin{align}
    \hat{c}_{\omega l}' &= \hat{c}_{\omega l}  + g (\omega) \cosh (r_\omega) \Big[ \varphi_\mathrm{CS} \big( \hat{c}_{\omega'r}^{\dd} - \hat{d}_{\omega'r} \big) + \alpha + \beta \Big],
    \vt 
\\
    \hat{d}_{\omega l}' &= \hat{d}_{\omega l}- g(\omega) \sinh (r_\omega)  \Big[ \varphi_\mathrm{CS} \big( \hat{c}_{\omega'r} - \hat{d}_{\omega'r}^{\dd} \big) + \alpha^\star + \beta^\star  \Big].
\vt 
\end{align}
We can conveniently express these as
\begin{align}
    \hat{c}_{\omega l}' &= \alpha g(\omega) \cosh (r_\omega)  + \hat{c}_{\omega l}'' ,
\vt 
\\
    \hat{d}_{\omega l}' &= - \alpha^\star g(\omega) \sinh (r_\omega) + \hat{d}_{\omega l}'' 
\vt 
\end{align}
where $c_\omega''$ and $d_\omega''$ are terms not multiplied by $|\alpha|$. The Unruh particle numbers can thus be calculated using
\begin{align}
    c_{\omega l}^{\dd \prime } c_{\omega l}' &=  |\alpha|^2 g^2(\omega) \cosh^2 (r_\omega)  + |\alpha|g(\omega)  \cosh (r_\omega) \left[ e^{i\phi} c_{\omega l}^{\dd\prime\prime} + e^{-i\phi}  c_{\omega l}''\right],
\vt 
\\
    d_{\omega l}^{\dd\prime} d_{\omega l}' &= |\alpha|^2 g^2(\omega) \sinh^2 (r_\omega )- |\alpha| g(\omega)  \sinh (r_\omega) \left[ e^{-i\phi}d_{\omega l}^{\dd\prime\prime} + e^{i\phi} d_{\omega l}'' \right] ,
\end{align}
where we have neglected terms not multiplied by $|\alpha| \gg 1$. The vacuum expectation value of the particle number is thus
\begin{align}
    \langle 0 | \hat{N} | 0 \rangle &= |\alpha|^2 \big( \mathcal{I}_\mathrm{C} + \mathcal{I}_\mathrm{S} \big) .
\end{align}
The relevant expectation values in $\langle 0 |\hat{N}^2|0 \rangle$ are
\begin{align}
\begin{split}
    \langle 0| c_{\omega l}'' c_{\omega' l}^{\dd\prime\prime}|0 \rangle &= \delta(\omega - \omega') + g(\omega) g(\omega') \cosh (r_{\omega}) \cosh (r_{\omega'}) \mathcal{I}_\mathrm{CS} ,
    \vt  
    \\
    \langle  0| c_{\omega l}^{\dd\prime\prime} c_{\omega' l}'' |0 \rangle &= g(\omega)g(\omega') \cosh (r_{\omega}) \cosh (r_{\omega'}) \mathcal{I}_\mathrm{CS}  ,
    \vt 
    \\
    \langle  0|d_{\omega l}'' d_{\omega' l}^{\dd\prime\prime} |0\rangle &= \delta(\omega - \omega') + g(\omega) g(\omega') \sinh (r_{\omega}) \sinh (r_{\omega'}) \mathcal{I}_\mathrm{CS}  
    \vt , 
    \\
   \langle  0| d_{\omega l}^{\dd\prime\prime}d_{\omega' l}'' |0\rangle &= g(\omega) g(\omega') \sinh (r_{\omega}) \sinh (r_{\omega'}) 
   \mathcal{I}_\mathrm{CS} , 
   \vt 
   \\
    \langle  0|c_{\omega l}'' d_{\omega' l}''  |0\rangle &= -g(\omega) g(\omega') \cosh( r_{\omega}) \sinh (r_{\omega'}) 
    \mathcal{I}_\mathrm{CS}  , 
    \vt 
   \\
    \langle  0|c_{\omega l}^{\dd\prime\prime}d_{\omega' l}^{\dd\prime\prime} |0\rangle &= -g(\omega) g(\omega') \cosh (r_{\omega}) 
    \sinh (r_{\omega'})
    \mathcal{I}_\mathrm{CS} 
    \vt 
    ,
    \\
    \langle  0| d_{\omega l}'' c_{\omega' l}''|0 \rangle &= - g(\omega) g(\omega') \sinh (r_{\omega}) \cosh (r_{\omega'}) \mathcal{I}_\mathrm{CS} , 
    \vt 
    \\
    \langle  0| d_{\omega l}^{\dd\prime\prime} c_{\omega' l}^{\dd\prime\prime} |0\rangle &= - g(\omega) g(\omega') \sinh (r_{\omega})
    \cosh (r_{\omega'})
    \mathcal{I}_\mathrm{CS} ,
    \vt 
\end{split}
\end{align}
where $\mathcal{I}_\mathrm{CS} = \int\D\omega \: g^2(\omega) \big( \cosh (r_{\omega}) - \sinh (r_{\omega}) \big)^2$. The terms in $\langle 0|\hat{N}^2|0\rangle$ are
\begin{align}
\begin{split}
    \langle 0 | \hat{c}_{\omega l}^{\dd\prime} \hat{c}_{\omega l}'\hat{c}_{\omega' l}^{\dd\prime} \hat{c}_{\omega' l}'|0\rangle &= |\alpha|^2 g^2(\omega) g^2(\omega') \cosh r_{\omega } \cosh r_{\omega' }  \left[ \delta(\omega - \omega') + 2\mathcal{I}_\mathrm{CS}' \cosh r_{\omega } \cosh r_{\omega' } \right]  
,
\vt 
\\
    \langle 0 | \hat{d}_{\omega l}^{\dd\prime}\hat{d}_{\omega l}'\hat{d}_{\omega' l}^{\dd\prime}  \hat{d}_{\omega' l}'|0\rangle &= |\alpha|^2 g^2(\omega) g^2(\omega') \sinh r_{\omega } \sinh r_{\omega' } \left[ \delta(\omega - \omega') + 2\mathcal{I}_\mathrm{CS}' \sinh r_{\omega } \sinh r_{\omega' } \right]
,
\vt 
    \\
    \langle 0 | \hat{c}_{\omega l}^{\dd\prime}\hat{c}_{\omega l}'\hat{d}_{\omega' l}^{\dd\prime} \hat{d}_{\omega' l}'|0\rangle &= 2|\alpha|^2 g^2(\omega) g^2(\omega') \cosh^2 r_{\omega } \sinh^2 r_{\omega' } \mathcal{I}_\mathrm{CS}' 
,
\vt 
\\
    \langle 0 | \hat{d}_{\omega l}^{\dd\prime}\hat{d}_{\omega l}'\hat{c}_{\omega' l}^{\dd\prime} \hat{c}_{\omega' l}'|0\rangle &= 2|\alpha|^2 g^2(\omega) g^2(\omega') \sinh^2 r_{\omega } \cosh^2 r_{\omega' } \mathcal{I}_\mathrm{CS}' ,
\vt 
\end{split}
\end{align}
where we have left out the terms fourth order in $|\alpha|$, since they are subtracted away by the $\langle 0 | \hat{N} | 0 \rangle^2$ term in the variance. Adding these terms together, integrating with respect to $\omega, \omega'$ and normalising by the strength of the local oscillator $|\alpha|^2 ( \mathcal{I}_\mathrm{C} + \mathcal{I}_\mathrm{S})$ yields Eq.\ (\ref{eqn:allopticalquadvar}). 

\section{Rindler-Squeezed Minkowski Vacuum State}
\noindent Like the displaced thermal state, we derived expressions for the output Unruh operators, given by
\begin{align}    
    \hat{c}_{\omega l}' &= \hat{c}_{\omega l} + g(\omega) \cosh r_{\omega } \Big[ (\cosh r - 1)  \int\D \omega'\:g(\omega') \big( \cosh r_{\omega'} \hat{c}_{\omega'l} + \sinh r_{\omega'} \hat{d}_{\omega'l}^{\dd} \big)\nonumber \\
    & + \sinh r  \int\D \omega'g(\omega') \big( \cosh r_{\omega'} \hat{c}_{\omega'l}^{\dd} + \sinh r_{\omega'} \hat{d}_{\omega'l} \big)  + \varphi_\mathrm{CS}( \hat{c}_{\omega'r}^{\dd} - \hat{d}_{\omega'r}\big) + \alpha \Big] \\
    \hat{d}_{\omega l}'&= \hat{d}_{\omega l} - g(\omega) \sinh r_{\omega } \Big[ (\cosh r - 1)  \int\D\omega'g(\omega') \big( \cosh r_{\omega'} \hat{c}_{\omega'l}^{\dd} + \sinh r_{\omega'} \hat{d}_{\omega'l} \big)  \nonumber \\
    & + \sinh r \int\D\omega'g(\omega') \big( \cosh r_{\omega'} \hat{c}_{\omega'l} + \sinh r_{\omega'} \hat{d}_{\omega'l}^{\dd} \big)  + \varphi_\mathrm{CS} \big( \hat{c}_{\omega'r}- \hat{d}_{\omega'r}^{\dd} \big) + \alpha^\star \Big] .
\end{align}
$\langle 0|\hat{N}|0\rangle$ is the same as previously derived. As before, the relevant expectation value for the products of Unruh operators are
\begin{align}
\begin{split}
    \langle 0| c_{\omega l}'' c_{\omega' l}'' |0\rangle &= g(\omega) g(\omega') \cosh r_{\omega}\cosh r_{\omega'} \psi_\mathrm{CC}  ,
    \vt  
    \\
    \langle 0| c_{\omega l}^{\dd\prime\prime}  c_{\omega' l}^{\dd\prime\prime}  |0\rangle &= g(\omega) g(\omega') \cosh r_{\omega}\cosh r_{\omega'}\psi_\mathrm{CC}   ,
    \vt  
    \\
    \langle 0| c_{\omega l}'' c_{\omega' l}^{\dd\prime\prime}  |0\rangle &= \delta(\omega - \omega') + g(\omega) g(\omega') \cosh r_{\omega}\cosh r_{\omega'} \bar{\phi}_\mathrm{CC}  ,
    \vt  
    \\
    \langle 0| c_{\omega l}^{\dd\prime\prime}  c_{\omega' l}''  |0\rangle &= g(\omega) g(\omega') \cosh r_{\omega}\cosh r_{\omega'} \phi_\mathrm{CC} ,
    \vt   
    \\
     \langle 0| d_{\omega l}'' d_{\omega' l}'' |0\rangle &= g(\omega) g(\omega') \sinh r_{\omega} \sinh r_{\omega'} \psi_\mathrm{DD} ,
     \vt  
     \\ 
    \langle 0| d_{\omega l}^{\dd\prime\prime} d_{\omega' l}^{\dd\prime\prime}  |0\rangle &= g(\omega) g(\omega') \sinh r_{\omega}\sinh r_{\omega'}\psi_\mathrm{DD} ,
    \vt  
    \\
    \langle 0| d_{\omega l}'' d_{\omega' l}^{\dd\prime\prime}  |0\rangle &= \delta(\omega - \omega') + g(\omega) g(\omega') \sinh r_{\omega}\sinh r_{\omega'} \bar{\phi}_\mathrm{DD} ,
    \vt  
    \\
    \langle 0| d_{\omega l}^{\dd\prime\prime} d_{\omega' l}'' |0\rangle &= g(\omega) g(\omega') \sinh r_{\omega}\sinh r_{\omega'}\phi_\mathrm{DD} ,
    \vt  
\end{split}
\end{align}
where
\begin{align}
\begin{split}
    \psi_\mathrm{CC}  &= \sinh r \big[  (\cosh r - 1) ( \mathcal{I}_\mathrm{C} + \mathcal{I}_\mathrm{S}) + 1\big]  ,
    \vt  
    \\
    \phi_\mathrm{CC} &= (\cosh r - 1)^2 \mathcal{I}_\mathrm{S} + \sinh^2 r \mathcal{I}_\mathrm{C} + \mathcal{I}_\mathrm{CS} ,
    \vt  
    \\
    \bar{\phi}_\mathrm{CC} &= 2(\cosh r -1 ) +  ( \cosh r - 1)^2 \mathcal{I}_\mathrm{C} + \sinh^2 r  \mathcal{I}_\mathrm{S} + \mathcal{I}_\mathrm{CS}   ,
    \vt  
    \\
    \psi_\mathrm{DD} &= \sinh r \big[ (\cosh r - 1) ( \mathcal{I}_\mathrm{C} + \mathcal{I}_\mathrm{S} )  - 1\big] ,
    \vt  
    \\
    \phi_\mathrm{DD} &= (\cosh r -1 )^2 \mathcal{I}_\mathrm{C} + \sinh^2 r \mathcal{I}_\mathrm{S} + \mathcal{I}_\mathrm{CS} ,
    \vt  
    \\
    \bar{\phi}_\mathrm{DD} &= -2 (\cosh r - 1) + ( \cosh r - 1)^2 \mathcal{I}_\mathrm{S} + \sinh^2r \mathcal{I}_\mathrm{C}  + \mathcal{I}_\mathrm{CS} .
    \vt  
\end{split}
\end{align}
For the cross-terms, we have
\begin{align}
\begin{split}
    \langle 0| c_{\omega l}'' d_{\omega' l}'' |0\rangle &= - g(\omega) g(\omega') \cosh r_{\omega} \sinh r_{\omega'} \big[ \bar{\phi}_\mathrm{CC} - (\cosh r - 1) \big]  
    ,
    \vt  
    \\
    \langle 0| c_{\omega l}^{\dd\prime\prime} d_{\omega' l}^{\dd\prime\prime}|0\rangle &= -g(\omega) g (\omega') \cosh r_{\omega} \sinh r_{\omega'}\big[ \bar{\phi}_\mathrm{DD} + (\cosh r - 1) \big] ,
    \vt  
    \\
    \langle 0| d_{\omega l}'' c_{\omega' l}'' |0\rangle &= -g(\omega) g (\omega') \sinh r_{\omega} \cosh r_{\omega'} \big[ \bar{\phi}_\mathrm{DD} + (\cosh r - 1) \big] 
    ,
    \vt  
    \\
    \langle 0| d_{\omega l}^{\dd\prime\prime} c_{\omega' l}^{\dd\prime\prime} |0\rangle &= -g (\omega) g(\omega') \sinh r_{\omega} \cosh r_{\omega'}\big[ \bar{\phi}_\mathrm{CC} - ( \cosh r - 1) \big]  ,
    \vt  
    \\
    \langle 0| c_{\omega l}'' d_{\omega' l}^{\dd\prime\prime} |0\rangle &= -g (\omega) g (\omega') \cosh r_{\omega} \sinh r_{\omega'} \gamma_\mathrm{CD} ,
    \vt  
    \\ 
    \langle 0| c_{\omega l}^{\dd\prime\prime} d_{\omega' l}'' |0\rangle &= -g(\omega) g(\omega') \cosh r_{\omega} \sinh r_{\omega'}\gamma_\mathrm{CD} ,
    \vt  
    \\
    \langle 0| d_{\omega l}'' c_{\omega' l}^{\dd\prime\prime}|0\rangle &= - g(\omega) g(\omega') \sinh r_{\omega} \cosh r_{\omega'}\gamma_\mathrm{CD}  ,
    \vt  
    \\
    \langle 0| d_{\omega l}^{\dd\prime\prime} c_{\omega' l}'' |0\rangle &= - g (\omega) g(\omega') \sinh r_{\omega} \cosh r_{\omega'}\gamma_\mathrm{CD} 
    \vt  
\end{split} 
\end{align}
where $\gamma_\mathrm{CD} = (\cosh r - 1) \sinh r ( \mathcal{I}_\mathrm{C} + \mathcal{I}_\mathrm{S})$. Thus, the relevant terms in $\langle 0|\hat{N}^2|0\rangle$ are
\begin{align}
    \langle 0 | \hat{c}_{\omega l}^{\dd\prime}\hat{c}_{\omega l}'\hat{c}_{\omega' l}^{\dd\prime} \hat{c}_{\omega' l}'|0\rangle &= |\alpha|^2 g^2(\omega) g^2(\omega') \cosh r_{\omega } \cosh r_{\omega' }  \big[ \delta(\omega - \omega') + 2 \bar{\phi}_\mathrm{CC} \cosh r_{\omega } \cosh r_{\omega' } + 2 \cos(2\phi) \cosh r_{\omega } \cosh r_{\omega' } \psi_\mathrm{CC} \big]  ,
    \nonumber  
    \vt  
    \\ 
    \langle 0 | \hat{d}_{\omega l}^{\dd\prime}\hat{d}_{\omega l}'\hat{d}_{\omega' l}^{\dd\prime}  \hat{d}_{\omega' l}'|0\rangle &= |\alpha|^2 g^2(\omega) g^2(\omega') \sinh r_{\omega } \sinh r_{\omega' }  \big[ \delta(\omega - \omega') + 2 \bar{\phi}_\mathrm{DD} \sinh r_{\omega l} \sinh r_{\omega' } + 2 \cos(2\phi) \sinh r_{\omega } \sinh r_{\omega' } \psi_\mathrm{DD} \big] \nonumber ,
    \vt  
    \\
    \langle 0 | \hat{c}_{\omega l}^{\dd\prime}\hat{c}_{\omega l}'\hat{d}_{\omega' l}^{\dd\prime} \hat{d}_{\omega' l}'|0\rangle &= |\alpha|^2 g^2(\omega) g^2(\omega') \cosh^2 r_{\omega } \sinh^2 r_{\omega' }  \big[ \bar{\phi}_\mathrm{CC} + \bar{\phi}_\mathrm{DD} + 2 \cos(2\phi) \gamma_\mathrm{CD} \big] \nonumber ,
    \vt  
    \\
    \langle 0 | \hat{d}_{\omega l}^{\dd\prime}\hat{d}_{\omega l}'\hat{c}_{\omega' l}^{\dd\prime} \hat{c}_{\omega' l}'|0\rangle &= |\alpha|^2 g^2(\omega) g^2(\omega') \sinh^2 r_{\omega} \cosh^2 r_{\omega' } \big[ \bar{\phi}_\mathrm{CC} + \bar{\phi}_\mathrm{DD} + 2 \cos(2\phi) \gamma_\mathrm{CD}\big] .
    \vt  
\end{align}
As before, we add these terms together and integrate over $\omega, \omega'$ to obtain the photon number variance. After normalising by the strength of the local oscillator, we obtain Eq.\ (\ref{eqn:squeezingvar}).
\\\\
\end{widetext}

\bibliography{Paper.bib}

\end{document}